%% file: main.tex
\documentclass[11pt]{article}
\usepackage{custom}
\usepackage{graphicx}
\usepackage{url}

\title{\textbf{Long-Baseline Atom Interferometry}}
\notoc

\begin{document}
\include{authorlist}
\maketitle

\section*{Abstract}
Long-baseline atom interferometry is a promising technique for probing various aspects of fundamental physics, astrophysics and cosmology, including searches for ultralight dark matter (ULDM) and for gravitational waves (GWs) in the frequency range around 1~Hz that is not covered by present and planned detectors using laser interferometry. The MAGIS detector is under construction at Fermilab, as is the MIGA detector in France. The PX46 access shaft to the LHC has been identified as a very suitable site for an atom interferometer of height around $100$ m, sites at the Boulby mine in the UK and the Canfranc Laboratory are also under investigation, and possible sites for km-class detectors have been suggested. The Terrestrial Very-Long-Baseline Atom Interferometry (TVLBAI) Proto-Collaboration proposes a coordinated programme of interferometers of increasing baselines.\\

{\centering
{\it Submission to the 2026 update of the European Strategy for Particle Physics on behalf of the TVLBAI Proto-Collaboration}\\
~~\\}

\noindent
%$^1$ Imperial College London,
%$^2$ CERN,
%$^3$ King's College London}
%$^4$  %https://indico.cern.ch/event/1369392/attachments/2789312/5096609/TVLBAI{\%}20Study{\%}20MOU{\%}20{\%}20Final{\%}20.pdf

\newpage

\section{Introduction}
The Terrestrial Very-Long-Baseline Atom Interferometry (TVLBAI) Proto-Collaboration~\cite{TVLBAIMOU} is coordinating international efforts to promote atom interferometers with baselines of $100$~m or more. These detectors aim to bridge significant gaps in our understanding of dark matter, gravitational waves (GWs) and fundamental physics using innovative quantum sensor technology~\cite{Buchmueller2023}. It is envisioned that they will operate as a network similar to the current LIGO, Virgo and KAGRA laser interferometers. In contrast to these detectors, which are most sensitive to GWs with frequencies $\sim 100$~Hz~\cite{Aasi2015,Acernese2014,Akutsu2020}, the proposed Einstein Telescope (ET)~\cite{Maggiore2020} and Cosmic Explorer (CE)~\cite{Reitze2019} laser interferometers that would be sensitive to frequencies above $10$~Hz and the planned LISA space-borne laser interferometer that will be most sensitive to GWs with frequencies $\sim 10^{-2}$~Hz~\cite{AmaroSeoane2017}, long-baseline atom interferometers will be most sensitive to GWs with intermediate frequencies $\sim 10^{-1} - 1$~Hz~\cite{Badurina2020,ElNeaj2020,Abe2021,Canuel2022}. Thus atom interferometers will complement the existing programme of GW detection with present and future laser interferometry experiments, and also probe fundamental physics and cosmology in novel ways. For example, long-baseline atom interferometers will search for possible interactions of bosonic ultralight dark matter (ULDM) with atomic constituents in a mass range beyond the reach of other current and planned experiments, and can provide sensitive probes of the equivalence principle. Many of the physicists working on atom interferometers have backgrounds in particle physics, and particle physics infrastructures such as Fermilab~\cite{Abe2021}, CERN~\cite{Arduini2023}, the Boulby mine in the UK and SURF in the USA are prospective sites for long-baseline atom interferometers. Further information about the TVLBAI project and its scientific goals are given in~\cite{TVLBAISummary} and \cite{Proceedings:2024foy} (provided as a supporting document), and the Memorandum of Understanding of the TVLBAI Proto-Collaboration is available from~\cite{TVLBAIMOU}.

\section{Science Goals}

The primary scientific objectives of the TVLBAI community have been discussed in two international workshops that each had over 200 participants drawn from the particle physics, atomic physics, astrophysics and cosmology communities~\cite{TVLBAISummary, Proceedings:2024foy}. Highlights of the proposed scientific programme were mentioned in the Introduction, and we present below a high-level summary of some primary science goals. 

\begin{figure}
\centering 
\includegraphics[width=0.4\textwidth]{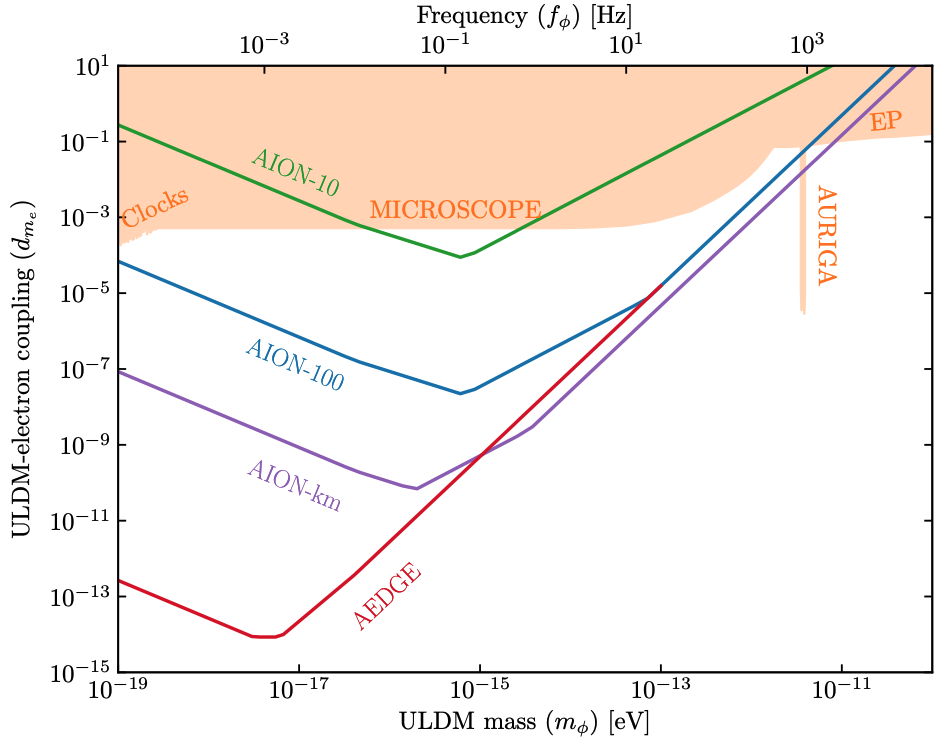}
\includegraphics[width=0.5\textwidth]{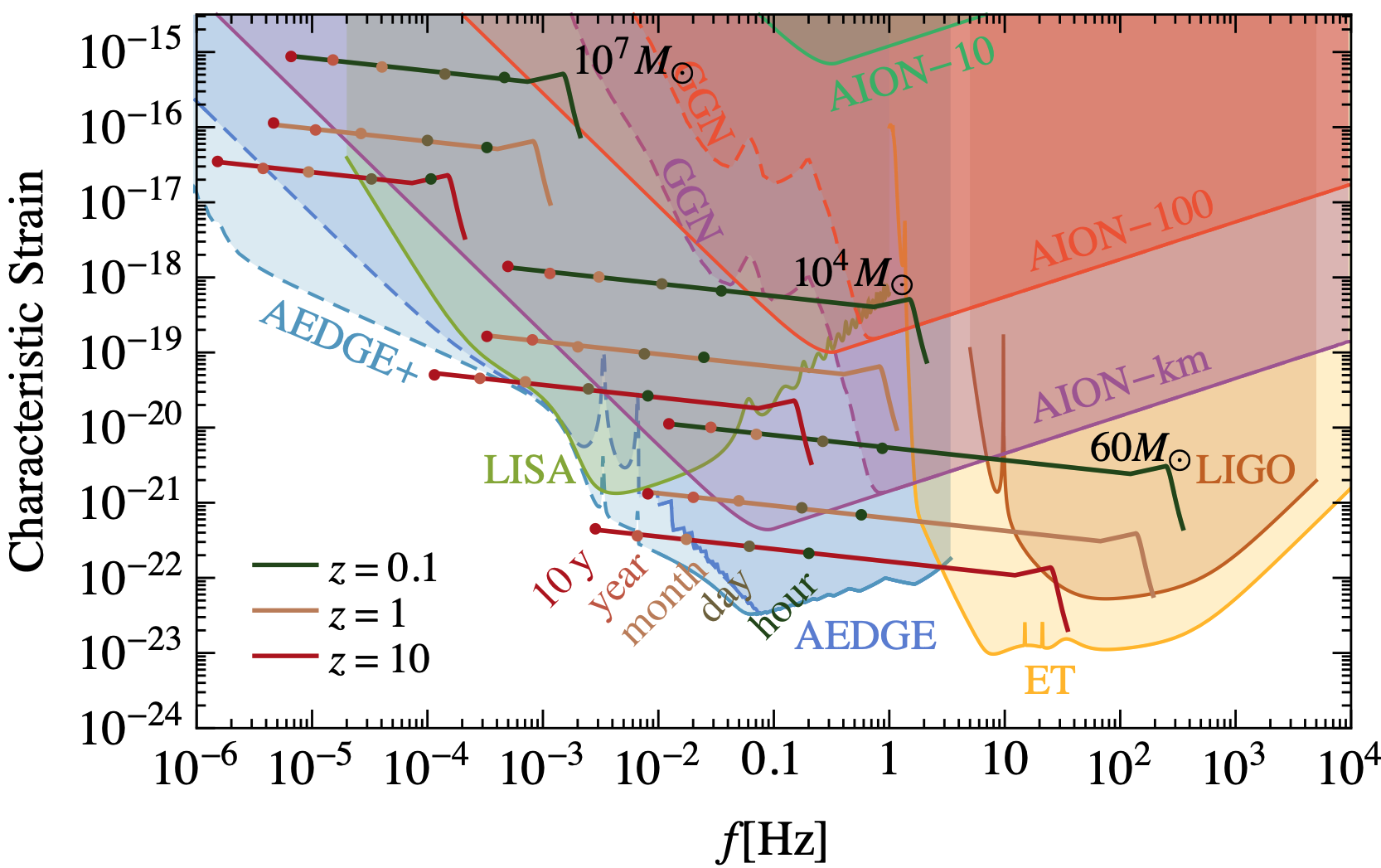}\\
\caption{\it Left panel: Projections for sensitivities to scalar ULDM linearly coupled to electrons. The shaded orange region is excluded by the existing constraints from searches for violations of the equivalence principle by the MICROSCOPE experiment and with torsion balances, atomic clocks, and the AURIGA experiment, as described in~\cite{Buchmueller2023}, where the assumed experimental specifications can be found. Right panel: The GW strain sensitivities and benchmark signals from BH binaries of different masses at various redshifts. The coloured dots indicate the times before mergers at which inspirals could be measured. Plot taken from~\cite{Badurina2021}.}
\label{fig:physics}
\end{figure}

\begin{itemize}
  
   \item \textbf{Dark Matter Exploration:} Astrophysicists and cosmologists infer the existence of dark matter from its gravitational effects, but there is no evidence that it is composed of particles, nor whether they have any other interactions. A very wide range of possible particle masses is allowed, and there are intensive searches for massive dark matter particles at accelerators and in scattering experiments underground, as well as many searches for lighter dark matter candidates such as axions~\cite{Chadha-Day:2021szb}. Atom interferometers would be sensitive to the possible interactions of waves of bosonic ultralight dark matter with atomic constituents~\cite{Arvanitaki:2016fyj, Badurina2020,ElNeaj2020}, and hence would be complementary to these other dark matter search experiments. Any signal would provide direct evidence of dark matter through non-gravitational interactions.

\item \textbf{Gravitational Wave Detection:} Long-baseline atom interferometers will target gravitational waves in the range of frequencies $\sim 0.1 - 10$~Hz, intermediate between the frequencies measured by LIGO/Virgo/KAGRA and those targeted by LISA, and outside the ranges accessible to ET and CE~\cite{Graham2013}. The capabilities of atom interferometers will fill a crucial observational gap, potentially unveiling novel phenomena. These include the mergers of intermediate-mass black holes with masses between 100 and $10^5$ solar masses that are thought to play important roles in the assembly of the supermassive black holes observed in the cores of galaxies~\cite{Ellis:2023iyb}. Moreover, networking atom interferometers with laser interferometers will enable synergistic observations of the evolution of specific sources over a wide range of frequencies, testing detailed general relativity calculations and setting tight constraints on the possible mass of the graviton~\cite{Ellis2020}. Measurements in the intermediate-frequency range may also reveal exotic compact objects that are predicted in some theories of physics beyond the Standard Model and conventional general relativity, but are currently undetectable~\cite{Banks2023}, or a stochastic background of gravitational waves generated by a first-order phase transition in the early Universe~\cite{Badurina2021}.
  
  \item \textbf{Fundamental Physics Tests:} Atom interferometer searches for dark matter of the type described above would also yield very sensitive probes of the Principle of Equivalence~\cite{Asenbaum2020}. They could also provide advanced tests of quantum mechanical phenomena such as the gravitational Aharonov-Bohm effect~\cite{Overstreet2022}, and explore fundamental interactions under conditions not accessible in conventional laboratory environments.
\end{itemize}

Figure~\ref{fig:physics} shows examples of the scientific reaches of TVLBAI searches for Ultra-Light Dark Matter (ULDM) and for GWs in the deciHz range~\cite{Buchmueller2023,Badurina2021}.

\section{TVLBAI Community}
Members of the TVLBAI Community have backgrounds in particle physics, atomic physics, astrophysics and cosmology, and are affiliated with major universities, national laboratories, and research institutes in Europe, North America and Asia. Thus they provide a wide range of expertise in fundamental physics, quantum technology, gravitational science and high-precision instrumentation. Members of the community are engaged in several TVLBAI precursor projects, including MIGA in France~\cite{Canuel2022}, VLBAI in Germany~\cite{schlippert2020matter}, ZAIGA in China~\cite{Zhan2019}, Stanford and MAGIS~\cite{Abe2021} in the USA and AION in the UK~\cite{Badurina2020}. 

During the second TVLBAI workshop~\cite{Proceedings:2024foy} the TVLBAI community held strategic discussions aimed at enhancing cooperation between institutions and coordinating plans for future TVLBAI projects. These discussions led to an initiative by the TVLBAI community to form a Proto-Collaboration governed by a Memorandum of Understanding (MoU) that has been signed by 53 institutions in 20 countries with 3 additional observer institutions~\cite{TVLBAIMOU}. 

\section{TVLBAI Proto-Collaboration Objectives}
 
 The objectives of the TVLBAI Proto-Collaboration are to reinforce collaboration between researchers tackling the challenges of designing and implementing km-scale atom interferometers. As outlined in its MoU, an important aspect of this effort is to develop a comprehensive roadmap that details design choices, technological considerations and science drivers for one or more kilometer-scale detectors that are expected to become operational in the mid-2030s. By pooling resources and expertise, the collaboration aims to advance the frontiers of atom interferometry and thereby enable new possibilities in fundamental physics research.

To this end, the TVLBAI Proto-Collaboration fosters the exchange of information between teams building precursor projects with baselines that are $\lesssim {\cal O}(100)$~m that are developing different technological approaches. These include the use of different atoms, such as rubidium, strontium and ytterbium, and different geometries, in either vertical shafts or horizontal galleries. The progress made within these different approaches will be exchanged and discussed within the TVLBAI Proto-Collaboration, and will inform the choices of technologies for km-scale detectors. The precursor detectors will also share measurements of the seismic and other environmental conditions at different sites, which will inform the choices of sites for km-class detectors.

The TVLBAI Proto-Collaboration will also provide a framework for networking between precursor projects, similar to that between the current LIGO, Virgo and KAGRA detectors, which will serve as a template for the networking of future km-scale detectors. The TVLBAI Proto-Collaboration will also encourage networking between the atom and laser interferometer communities, which is expected to yield synergies between detectors making observations in different frequency ranges, These could, for example, enable GW measurements over the full history of a merger, including inspiral, infall and ringdown as well the actual merger itself.

As part of its framework for the exchange of information and coordination between long-baseline atomic interferometry projects, the TVLBAI Proto-Collaboration has set up international topical working groups. It is also scheduling regular in-person meetings, previously at CERN and Imperial College London, in the future in Hanover and at the Canfranc laboratory. Equity, diversity, and inclusion are enshrined in the TVLBAI MoU~\cite{TVLBAIMOU}.

%\section{Collaboration Composition}
%The TVLBAI Collaboration consists of a diverse and international consortium of leading researchers and institutions. Partners include major universities, national laboratories, and research institutes from Europe, North America, and Asia, ensuring a wide range of expertise in quantum physics, gravitational science, and high-precision instrumentation. 

\section{Facility Requirements for Long-Baseline Atom Interferometers}

As already mentioned, two possible geometrical configurations are under consideration for TVLBAI projects: in a vertical shaft or a horizontal gallery. There are 10m prototype vertical devices at Stanford University and in Hanover~\cite{schlippert2020matter}. The MAGIS Collaboration is currently constructing a 100m detector in an access shaft of the Fermilab neutrino beam~\cite{Abe2021}, the MIGA Collaboration is currently constructing a two-arm horizontal detector at an underground laboratory in France~\cite{Canuel2022}, and the ZAIGA programme is underway in China~\cite{Zhan2019}. Supported by the CERN Physics Beyond Colliders study group, the PX46 access shaft to the LHC has been identified as a very suitable site for a vertical atom interferometer of length ${\cal O}(100)$m~\cite{Arduini2023}, which would align with the CERN Quantum Technology Initiative, and the Boulby laboratory in the UK and the Canfranc laboratory in Spain have also been proposed as possible sites for vertical detectors with similar baselines~\cite{Proceedings:2024foy}.

ELGAR is a project for a multi-km two-arm horizontal detector~\cite{Canuel2020}, and several sites have been proposed for possible km-scale vertical detectors, including Boulby in the UK, SURF in the US and Porta Alpina in Switzerland~\cite{Proceedings:2024foy}. Hosting a vertical long baseline detector would require a facility meeting a specific set of infrastructure and environmental conditions that are critical to ensure the sensitivity and accuracy of our experiments, which are summarized below. 

\subsection{Infrastructure Requirements}

\begin{table}
\centering
\begin{tabular}{|l|l|}
\hline
{\bf Component}       & \multicolumn{1}{c|}{{\bf Requirements}}                    \\
\hline
Laser laboratory      & At least 50 m$^2$ area                                     \\
                      & 35 kW electrical power                                     \\
                      & Air cooling: 30 kV heat load, 1 $^\circ$C stability        \\
                      & Maximum distance to interferometry region: 50 m            \\
\hline
Interferometry region & Tunnel or shaft with required baseline length              \\
                      & Full access to entire tunnel or shaft                      \\
\hline
Atom sources          & 100~m: 2 to 10 units, spaced over the tunnel or shaft     \\
& 1~km: Up to 100 units, spaced over the tunnel or shaft     \\
                      & 1$\times$1$\times$2 m$^3$ volume, 200 kg weight (per unit) \\
                      & 10 kW power consumption (per unit)                         \\
\hline
\end{tabular}
\caption{Infrastructure requirements for TVLBAI experiments. Adapted from~\cite{Arduini2023}.}
\label{table-sites-requirements}
\end{table}

The infrastructure to support a vertical TVLBAI experimental setup is summarised in Table~\ref{table-sites-requirements} and includes:
\begin{itemize}
  \item \textbf{Shaft Requirements:} The shaft should have a minimum diameter of approximately 1.5 meters to accommodate the full setup of the experiment including access for construction, commissioning and maintenance. The shaft must provide a baseline with an unimpeded line of sight to ensure the integrity and feasibility of experimental measurements.
  \item \textbf{Interferometry Region:} A vertical tube equipped with ultra-high vacuum conditions across the entire baseline, and a minimum diameter of 15~cm is required to house the experimental apparatus.
  \item \textbf{Laser Laboratory:} A dedicated space of at least 50 m$^2$, equipped with power supplies capable of providing 35 kW, and stringent temperature control to maintain fluctuations below 1ºC, which might be located at either end of the tube.
  \item \textbf{Atom Sources:} Facilities for atom sources that provide at least 2 m$^3$ per unit, can support weights up to 200 kg, and offer 10 kW of electrical power for each unit.
\end{itemize}

\subsection{Environmental Requirements}
To ensure the precision of the measurements, the chosen site should have:
\begin{itemize}
  \item \textbf{Minimal Gravity Gradient Noise (GGN):} This is essential for reducing sensitivity degradation due to environmental stochastic matter perturbations. Seismic studies have been conducted at the MAGIS site~\cite{Abe2021} and at the PX46 LHC access shaft~\cite{Arduini2023}, which have found seismic noise levels that are intermediate between those in the new high- and low-noise models discussed in~\cite{Peterson1993}.
  \item \textbf{Low Electromagnetic Noise:} This is critical to prevent interference with the delicate instrumentation of the interferometer. Measurements at the PX46 LHC access shaft~\cite{Arduini2023} have established that the electromagnetic noise level is sufficiently low at all times, including during LHC operations.
\end{itemize}

\section{Tentative Timeline}
\begin{itemize}
  \item \textbf{2025-2029:} Detailed design, prototyping, and initial ground tests at smaller-scale facilities. The MAGIS 100~m and MIGA detectors are under construction, and a 100~m detector could be installed in the PX46 shaft at CERN during LS3.
  \item \textbf{2029-2034:} Construction, installation, and commissioning of one or more full-scale interferometers.
  \item \textbf{2035 onwards:} Commencement of full operational phase with continuous data collection and analysis, aiming for significant scientific outputs and discoveries.
\end{itemize}

\section{Tentative Cost Estimates}
Using the estimated costs of current long-baseline atom interferometers as a basis, one may make tentative, preliminary estimates of the possible costs of very long baseline detectors and their precursors, assuming the availability of a refurbished shaft that is ready for construction. This is largely the case for the CERN PX46 option, apart from an estimated cost of about 1.5M CHF to isolate it from LHC and provide a mobile operational platform, access, safety and monitoring systems, general services and utilities~\cite{Arduini2023}. However, these costs could be quite different for a 100m shaft requiring more preparatory work, or for a km-scale detector.

A preliminary estimate for an initial 100m detector with a capability to search for ULDM is about 30M CHF, and an upgrade to reach the ultimate sensitivity is estimated to cost about 20M CHF. These figures include 25\% contingency allowances.

Extrapolating to the km-scale, a preliminary estimate of the cost is an order of magnitude greater.
%for an initial detector is about 250M to 300M CHF, with a cost for an upgrade option of an additional 150M to 200M CHF.

\section{Summary}

The TVLBAI Proto-Collaboration expresses its strong interest in advancing the frontiers of science through the unique capabilities of atom interferometers, which align well with various national quantum strategies. We give below explicit answers to the ESPP questions for proposed projects.

The proposed programme would proceed in the following stages:

\begin{itemize}
    \item A global network of detectors with baselines ${\cal O}(100)$~m, which would provide unique sensitivity to the possible couplings of bosonic ultra-light dark matter to atomic constituents - see the left panel of Fig.~\ref{fig:physics} - and a first exploration of possible gravitational wave signals with frequencies ${\cal O}(1)$~Hz - see the right panel of Fig.~\ref{fig:physics}. One of these detectors could be located at CERN in the PX46 access shaft to the the LHC, in the Boulby mine in the UK or at the Canfranc laboratory in Spain.
    \item One or more detectors with baselines ${\cal O}(1)$~km, which would extend the search for ultra-light dark matter couplings and offer the possibility of detecting and measuring gravitational waves produced by the mergers of intermediate-mass black holes. Possible sites under investigation include SURF in the US, an access shaft to the Gotthard Base Tunnel in Switzerland and the Boulby mine in the UK.
\end{itemize}

This logical sequence would follow on from existing and planned prototype detectors with baselines ${\cal O}(10)$~m located in national universities and laboratories.

The 100-m detectors could be built and operated by collaborations of a few dozen physicists and engineers from the particle and cold atom communities, whereas a collaboration numbered in the hundreds would be required for a km-scale detector.

Prospective timelines for the stages mentioned above are given in Section~6. We note that LS3 would provide an excellent opportunity to install a 100-m detector in PX46, which could be built and operated independently of LHC operations.

Tentative cost estimates for the two stages are given in Section~5: 30M to 50M CHF for a 100-m detector and an order of magnitude greater for a km-class detector. The supplementary cost of the necessary infrastructure work at CERN was estimated in~\cite{Arduini2023} to be about 1.5M CHF.

A complete life-cycle assessment will be undertaken during the design stage. Nevertheless, the environmental impact is expected to be small, as no substantial civil engineering would be required, and the power consumption would also be small: 35~kW for the laser system and at most 100~kW (1~MW) for the atom sources for a 100-m (1-km) detector.

The key cold atom technologies are currently being developed and deployed in prototype devices in France, Germany, Italy, the UK and the US.

As mentioned above, CERN is one possible site for a 100-m detector, which could be built and operated independently from LHC operations. Detectors of similar sizes are under construction at Fermilab (MAGIS) and in France (MIGA).

The current project status is summarised in~\cite{Abe2021,Canuel2022,Buchmueller2023, TVLBAISummary,Proceedings:2024foy}, and there is widespread interest in a global Proto-Collaboration~\cite{TVLBAIMOU} to coordinate long-baseline atom interferometer projects.

\newpage
%SURF. SURF's infrastructure, upcoming expansions, and the natural characteristics of the site position it ideally to meet the requirements of the TVLBAI project for a vertical baseline detector. These features, combined with SURF's dedication to supporting high-level scientific research, make it an exemplary candidate for establishing a vertical baseline atom interferometer. We believe our project not only aligns with the scientific goals of SURF, but also has the potential to contribute significantly to the global scientific community's understanding of the Universe. We look forward to the possibility of a collaborative and productive relationship with SURF.

\bibliographystyle{JHEP}
\bibliography{main}

\end{document}

%% file: authorlist.tex
\author[1,*]{Antun~Bala\v{z},} \affiliation[1]{Institute of Physics Belgrade, University of Belgrade, Pregrevica 118, 11080 Belgrade, Serbia}
\author[2,*]{Diego~Blas,} \affiliation[2]{Institut de F\'{i}sica d'Altes Energies (IFAE), The Barcelona Institute of Science and Technology, Campus UAB, 08193 Bellaterra (Barcelona), Spain;\\ Instituci\'{o} Catalana de Recerca i Estudis Avan\c{c}ats (ICREA), Passeig Llu\'{i}s Companys 23, 08010 Barcelona, Spain}
\author[3,*,@]{Oliver~Buchmueller,} \affiliation[3]{Imperial College London, Prince Consort Road, London, SW7 2AZ, UK}
\author[4,*]{Sergio~Calatroni,} \affiliation[4]{CERN, 1211 Geneva 23, CH}
\author[5,*]{Laurentiu-Ioan~Caramete,} \affiliation[5]{Institute of Space Science – INFLPR Subsidiary, Atomistilor Str., no. 409, Magurele, Ilfov, Romania, 077125, RO}
\author[6,*]{David~Cerde\~no,} \affiliation[6]{Instituto de F\' \i sica Te\'orica, IFT-UAM/CSIC, 28049 Madrid, Spain}
\author[7,*]{Maria Luisa Chiofalo,} \affiliation[7]{Department of Physics, University of Pisa, Largo Bruno Pontecorvo 3 56126 Pisa, Italy}
\author[8,*]{Fabio~Di~Pumpo,} \affiliation[8]{Institut f{\"u}r Quantenphysik, Universität Ulm, Albert-Einstein-Allee 11, D-89069 Ulm, Germany}
\author[9,*]{Goran~Djordjevic,} \affiliation[9]{SEENET-MTP Centre, Ilije Garasanina 45, Nis, 18000, Serbia}
\author[10,\%]{John~Ellis,} \affiliation[10]{King's College London, Strand, London, WC2R 2LS, UK}
\author[11,*]{Pierre~Fayet,} \affiliation[11]{Laboratoire de physique de l'ENS, Ecole normale sup\'erieure-PSL, CNRS, Sorbonne Universit\'e, Universit\'e Paris Cit\'e, 24 rue Lhomond, 75231 Paris Cedex 05, France}
\author[12,*]{Chris~Foot,} \affiliation[12]{University of Oxford, South Parks Road, Oxford OX1 3PU, UK}
\author[13,*]{Naceur~Gaaloul,} \affiliation[13]{Leibniz Universit\"at Hannover, Welfengarten 1, 30167 Hannover, Germany}
\author[14,*]{Susan~Gardner,} \affiliation[14]{Department of Physics and Astronomy, University of Kentucky, Lexington, KY 40506-0055, USA}
\author[15,*]{Barry~M~Garraway,} \affiliation[15]{Department of Physics \& Astronomy, University of Sussex, Falmer, Brighton, BN1 9QH, UK}
\author[16,*]{Alexandre~Gauguet,} \affiliation[16]{University of Toulouse, Toulouse, 118 Route de Narbonne, France}
\author[17,*]{Enno~Giese,} \affiliation[17]{TU Darmstadt, Schlossgartenstr. 7, 64289 Darmstadt, Germany}
\author[18]{Jason~M.~Hogan,} \affiliation[18]{Department of Physics, Stanford University, Stanford, California 94305, USA}
\author[19,*]{Onur~Hosten,} \affiliation[19]{Institute of Science and Technology Austria, Am Campus 1, 3400 Klosterneuburg, Austria}
\author[20,*]{Alex~Kehagias,} \affiliation[20]{National Technical University, Herron Polytechniou 9, Athens, 15780, GR}
\author[21,*]{Eva~Kilian,} \affiliation[21]{University College London, London WC1E 6BT, UK}
\author[22,*]{Tim~Kovachy,} \affiliation[22]{Northwestern University, Evanston, Illinois, USA}
\author[23,*]{Carlos~Lacasta,} \affiliation[23]{Instituto de Física Corpuscular- IFIC/CSIC-UV, Valencia, Spain}
\author[24,*]{Marek~Lewicki,} \affiliation[24]{ University of Warsaw, Faculty of Physics, Pasteura 5, 02-093 Warsaw, Poland }
\author[6,*]{Elias~Lopez~Asamar,} \affiliation[6]{Instituto de F\' \i sica Te\'orica, IFT-UAM/CSIC, 28049 Madrid, Spain}
\author[25,*]{J.Luis~Lopez-Gonzalez,} \affiliation[25]{Department of Mathematics and Physics, Autonomous University of Aguascalientes, Av. Universidad 940, Aguascalientes 20100, Mexico}
\author[26,*]{Nathan~Lundblad,} \affiliation[26]{Bates College, Lewiston, Maine, USA}
\author[27,*]{Michele~Maggiore,} \affiliation[27]{University of Geneva, Geneva, Switzerland}
\author[10,*]{Christopher~McCabe,} %\affiliation[10]{King's College London, Strand, London, WC2R 2LS, UK}
\author[28,*]{John~McFerran,} \affiliation[28]{University of Western Australia, 35 Stirling Highway, Crawley, 6009, Australia}
\author[29,*]{Gaetano~Mileti,} \affiliation[29]{University of Neuchâtel, Switzerland}
\author[30,*]{Peter~Millington,} \affiliation[30]{Department of Physics and Astronomy, University of Manchester, Manchester M13 9PL, UK}
\author[31,*]{Gavin~W.~Morley,} \affiliation[31]{Warwick University, Coventry, CV4 7AL, UK}
\author[32,*]{Senad~Od\v{z}ak,} \affiliation[32]{University of Sarajevo - Faculty of Science, Zmaja od Bosne 33-35, 71000 Sarajevo, Bosnia and Herzegovina}
\author[33,*]{Chris~Overstreet,} \affiliation[33]{The Johns Hopkins University, Baltimore, MD 21218, USA}
\author[34,*]{Krzysztof~Pawlowski,} \affiliation[34]{Center for Theoretical Physics PAS, Aleja Lotnikow 32/46, PL-02-668 Warsaw, Poland}
\author[35,*]{Emanuele~Pelucchi,} \affiliation[35]{Epitaxy and Physics of Nanostructures, Tyndall National Institute, University College Cork, Lee Maltings, Dyke Parade, Cork, T12R5CP, Ireland}
\author[36,*]{Johann Rafelski,} \affiliation[36]{Department of Physics, The University of Arizona, Tucson, AZ 85721, USA}
\author[37]{Albert~Roura,} \affiliation[37]{German Aerospace Center (DLR), 89081 Ulm, Germany}
\author[38,*]{Marianna S. Safronova,} \affiliation[38]{University of Delaware, Newark, Delaware 19716, USA}
\author[39,*]{Florian Schreck,} \affiliation[39]{University of Amsterdam, Science Park 904, 1098 XH Amsterdam, The Netherlands}
\author[40,*]{Olga Sergijenko,} \affiliation[40]{Space Technology Centre, AGH University of Krakow, Aleja Mickiewicza, 30, 30-059, Krakow, Poland}
\author[41,*]{Yeshpal~Singh,} \affiliation[41]{University of Birmingham, Edgbaston Road, Birmingham, B15 2TT, UK}
\author[42,*]{Marcelle~Soares-Santos,} \affiliation[42]{University of Zurich, Winterthurerstrasse 190, Zurich, 8057 Zurich, Switzerland}
\author[43,*]{Nikolaos Stergioulas,} \affiliation[43]{Department of Physics, Aristotle University of Thessaloniki, GR-54124 Thessaloniki, Greece}
\author[44,*]{Guglielmo M.\ Tino,} \affiliation[44]{Dipartimento di Fisica e Astronomia and LENS Laboratory, Universit\`a di Firenze, via  Sansone 1,  Sesto Fiorentino, Italy}
\author[45,*]{Jonathan N.~Tinsley,} \affiliation[45]{Department of Physics, University of Liverpool, Liverpool, Merseyside, UK}
\author[46]{Hendrik~Ulbricht,} \affiliation[46]{University of Southampton, Southampton, SO17 1BJ, UK}
\author[47,*]{Maurits~van der Grinten,} \affiliation[47]{Rutherford Appleton Laboratory, UKRI-STFC, Fermi Avenue, Didcot, OX11 OQX, UK}
\author[48,*]{Ville~Vaskonen,} \affiliation[48]{Keemilise ja Bioloogilise F\"u\"usika Instituut, R\"avala pst. 10, 10143 Tallinn, Estonia}
\author[49,*]{Wolf~von~Klitzing,} \affiliation[49]{Foundation for Research and Technology -  Hellas (FORTH), Institute of Electronic Structure and Lasers (IESL), Vassilika Vouton, 70013 Heraklion, Creta, Greece}
\author[50,*]{André~Xuereb,} \affiliation[50]{University of Malta, Msida, MSD2080, Malta}
\author[51,*]{Emmanuel~Zambrini Cruzeiro,} \affiliation[51]{Instituto de Telecomunicações, Av. Rovisco Pais 1, 1049-001, Lisboa, Portugal}
\affiliation[\%]{Main Editor: John.Ellis@cern.ch}
\affiliation[@]{Chair of the TVLBAI Proto-Collaboration: Oliver.Buchmueller@cern.ch}
\affiliation[*]{Member of the International Collaboration Board of the TVLBAI Proto-Collaboration, signing on behalf of their institution.}
%\emailAdd{Oliver.Buchmueller@cern.ch} %\emailAdd{John.Ellis@cern.ch} 